# Resilient Cloud cluster with DevSecOps security model, automates a data analysis, vulnerability search and risk calculation

**Abed Saif Ahmed Alghawli [1],[*] and Tamara Radivilova [2]**

[1] Department of Computer Science, College of Sciences and Humanities, Prince Sattam Bin Abdulaziz University, Aflaj, Kingdom of Saudi Arabia; a.alghauly@psau.edu.sa

[2] Department of Infocommunication Engineering, Kharkiv National University of Radio Electronics, Kharkiv, Ukraine; tamara.radivilova@nure.ua

[*] Correspondence: Abed Saif Ahmed Alghawli, a.alghauly@psau.edu.sa

**Abstract**

Automated, secure software development is an important task of digitalization, which is solved with the DevSecOps approach. An important part of the DevSecOps approach is continuous risk assessment, which is necessary to identify and evaluate risk factors. Combining the development cycle with continuous risk assessment creates synergies in software development and operation and minimizes vulnerabilities. The article presents the main methods of deploying web applications, ways to increase the level of information security at all stages of product development, compares different types of infrastructures and cloud computing providers, and analyzes modern tools used to automate processes. The cloud cluster was deployed using Terraform and the Jenkins pipeline, which is written in the Groovy programming language, which checks program code for vulnerabilities and allows you to fix violations at the earliest stages of developing secure web applications. The developed cluster implements the proposed algorithm for automated risk assessment based on the calculation (modeling) of threats and vulnerabilities of cloud infrastructure, which operates in real time, periodically collecting all information and adjusting the system in accordance with the risk and applied controls. The algorithm for calculating risk and losses is based on statistical data and the concept of the FAIR information risk assessment methodology. The risk value obtained using the proposed method is quantitative, which allows more efficient forecasting of information security costs in software development.

**1. Introduction**

Modern information technologies are developing at an ever-increasing rate and are present in almost every sphere of life. Cloud computing has now become the most advanced technology in the world, and two challenges are in the spotlight: how to optimize cloud computing resources and how to ensure information security when using cloud technologies if risks are to be effectively minimized [1]. Companies not only store their data in the clouds, but also use cloud technologies to develop and run software, web applications, etc. Software development usually uses cloud environments and virtualization in them [2] to accelerate development, but the pressure of deadlines often leads to neglect of information security important aspects. In this environment, it is difficult for development teams to constantly review and consistently secure their software code at all stages of development and operation, as demonstrated by the continuous increase the number of vulnerabilities that the non-profit MITRE Corporation publishes annually in its CVE (Common Vulnerabilities and Exposures) list. Regular security patches, even by well-known software vendors, are an indicator of the titanic work done to eliminate vulnerabilities, not to mention detecting them before they appear on the market. Thus, there is a contradiction between the requirements for software development and the requirements for program code security. This contradiction is being resolved by adherence to modern software development methods, as well as compliance with DevSecOps approaches, considering the risks of vulnerability exploitation and the introduction of cloud technologies into the activities of organizations [3, 4].

There are approaches to the automated calculation of vulnerabilities and risks of organizations when using cloud technologies, but there is no data on their implementation in cloud systems during software development. There are also difficulties in automating and quantitative risks [5, 6], as well as problems of lack of time for testing, implementing, and updating software [7, 8]. Existing approaches allow for separate automation of secure software development, separate calculation of vulnerabilities in program code, and separate calculation of risks to the organization's functioning. However, unlike existing approaches, solving the three tasks together makes it possible to automatically calculate a more accurate value of the risks of an organization's functioning and identify more software and organization vulnerabilities.

Based on the above, there is a need to develop methods that will ensure the reliability of the organization's cyber infrastructure, web application security, quality of service, and vulnerability detection at all stages of product development [9-13], and which should take into account the quantitative risk assessment and vulnerabilities of the organization's activities and the use of cloud technologies. Our proposed approaches make it possible to complexly resolve the above contradiction of automated fast, reliable, and secure software development with the ability to automatically detect vulnerabilities and quantitative risks in the program code and the organization's functioning.

The objective of this paper is to build a cloud cluster with automated calculation of vulnerabilities and quantitative risk assessment in secure software development. To realize this goal, the authors have solved the following tasks:
- to review existing scientific works to solve the objective (Section 2);
- methods for creating a secure infrastructure in the cloud environment are analyzed (Section 3);
- an algorithm for automated quantitative risk assessment based on the calculation (modeling) of threats and vulnerabilities in the organization and cloud infrastructure is proposed. Moreover, threat modeling for security risks focuses on all stages of the software development life cycle (Section 4);
- to calculate risks, an approach based on the modernized methodology of factor analysis of information risks (FAIR) is proposed, which provides a quantitative value of risk (Section 4);
- the method of analysis of hierarchies by Thomas Saaty is proposed to find the most relevant cloud provider for the organization, mathematical calculations, analysis, and results of choosing AWS provider as the basis for building a cloud cluster are presented (Section 5.1);
- in the process of building a cloud cluster, the authors have developed and provided recommendations for creating a fault-tolerant and secure infrastructure in the AWS cloud provider, as well as methods for improving information security when deploying and implementing web applications (Section 5.2);
- summarizing the proposed methods and solved problems (Section 6).

Achieving the goal and solving the tasks will allow companies to implement a cloud-based cluster for automated development of secure software throughout the entire life cycle, taking into account vulnerabilities and quantitative risk assessment of the company. This will help to increase the reliability and security of the organization's cyber infrastructure, and software security, improve the quality of service, and detect vulnerabilities at all stages of product development and organization functioning.

## 2. Related Research

To ensure secure software development, it is necessary to use the DevSecOps approach, but its implementation increases the time required to develop a product and put it into production. The necessity of using the DevSecOps approach and its advantages and disadvantages are analyzed in the following papers. After analyzing research, specifications, and various case studies, the authors of [14] listed a set of factors related to the key categories of DevSecOps and the difficulties of its implementation. Based on a questionnaire survey, the authors identified the following factors that are of greatest concern and impede the successful implementation of DevSecOps: lack of secure coding standards, lack of automated security testing tools in DevOps, lack of knowledge of static security testing, and failure to communicate the security standard to the DevOps team. The authors of [15] note that companies that have implemented DevSecOps focus on automation, security, and vulnerabilities when using the tools. However, there is a lack of tools for continuous assessment of vulnerabilities and security risks. To address these issues of developer training and continuous automated assessment of security vulnerabilities and risks, the authors of [16] developed a scalable automated platform for cybersecurity training in the clouds. The authors proposed a methodology for automating the modeling of malicious cyberattacks integrated into Ansible and based on the DevSecOps strategy and the principles of infrastructure as code (IaC). However, its implementation is hampered by the lack of time for developers to learn and the absence of constant vulnerability updates and risk calculations. In addition, these approaches do not address software development issues. In article [17], the authors investigated the DevSecOps approach in terms of software and application development, operation, and security. A background analysis and a survey of experts were conducted to understand DevSecOps, practices, and existing problems associated with the implementation of DevSecOps in software development. As a result of the analysis, three security risks and three aspects of DevSecOps were identified: people, processes, and technology selection. These studies point to the need to address the issue of processes and technology selection through automation using the DevSecOps approach, especially when developing software using cloud technologies. These issues were partially resolved by the authors of [18], who developed a self-service approach to cybersecurity monitoring during software deployment and testing. The practical implementation of the proposed approach was developed on a specific example of setting up a cybersecurity monitoring infrastructure using virtualization and containerization technologies. However, this study does not include a risk and vulnerability calculation.

The author of [19] proposes an approach to risk management throughout the entire life cycle of critical infrastructure systems development using DevSecOps technology. At the same time, the risk assessment is qualitative rather than quantitative, based on expert opinions and, accordingly, not automated. In [20], the authors implemented the DevSecOps approach to automate work in the organization and manage risks to enhance security, including people and technology. At the same time, the organization implemented quality control, version control, testing, and release of security-based software products. However, the risks were expert, taken from the company's database, and were constants, meaning that continuous risk and vulnerability assessment could not be performed. In [21], the authors conducted a benchmarking analysis to identify risk factors and the most influential functions in cloud computing. The authors used machine learning algorithms to analyze threats, and an expert survey was conducted to confirm the results. The results of the analysis showed that security risks are the most influential when using cloud computing. During the background analysis, the authors of [22] found that information security risks are the most significant and frequent. The most significant risk factors are data leakage, multi-user mode, choosing the right provider. The authors also described environmental, technological, and organizational factors of information security risks. The authors of the patent [23] propose a method for assessing risks for an enterprise using cloud services from one or more cloud service providers. This method includes generating risk assessments for the enterprise based on the provider's risk assessments, behavior and structure of cloud services used by the enterprise. However, in these studies, the risk assessment process is not quantitative and not automated.

In [24], the authors analyze risk assessment models and compare their strengths and weaknesses. Based on the analysis, the authors propose their own risk assessment model that adequately takes into account all the characteristics of cloud computing that were not taken into account in existing models. However, the risk assessment is expert and not automated, and the authors do not take into account system vulnerabilities and software development risks. The authors of [25] propose a hierarchical risk assessment model that they have implemented on a cloud platform and which allows the platform to independently assess its security status. The authors also evaluate the usability of the proposed model by simulating distributed denial-of-service and error injection attacks. The calculation is quantitative and automated, but also based on expert judgment and does not take into account software development risks. To take into account the vulnerabilities and risks of software development in cloud systems, approaches based on the use of risk management methodologies are used. In [26], the authors analyzed the most common risk management methodologies and their evaluation in terms of ease of use, adaptability, and inclusion in cloud systems. The authors found that the best models for cloud hosting are OCTAVE Allegro, COBIT 5, and CORAS. When assessing risks, these models include cloud infrastructure and ensure the confidentiality, availability and integrity of information assets. T. Wail in his work [27] described approaches to the risks of deploying Saas, IaaS and PaaS software in the cloud using the ISO 27001 standard. The author showed that this standard is only 10% of the Cloud Security Alliance's cloud governance matrix and gave examples of how this standard can be used to implement risk assessment for cloud computing platforms. The author also presented the three main risks in the cloud: IT failures, customer data protection, and regulatory compliance. But these works do not include the risks and vulnerabilities of software development.

The authors of [28] analyze the impact of network failures on cloud applications using the Factor Analysis of Information Risk (FAIR) model (discussed in [29-31]) and server failure data from the Panopta monitoring service. Using the Monte Carlo method, the authors quantify the risk of network failures using a geographic analysis, i.e., different server service providers in various regions. The authors also propose strategies to reduce the associated risks when implementing cloud technologies in enterprises. However, this approach is not automated and does not take into account system vulnerabilities in software development.

Thus, based on the analysis of existing studies, we can see that there are approaches to calculating risk in cloud systems, calculating risk in software development, but they are either not automated, or not quantitative, or do not take into account the vulnerabilities of using cloud technologies in software development, or are used only for one cloud provider. Thus, there is a need to develop an approach to automatically calculate the risks and vulnerabilities of using cloud systems in the development of secure software. Also, as noted in [22], it is necessary to develop an approach to choosing a cloud provider. Therefore, in our work, we have developed an approach to selecting a cloud provider, developed a cloud cluster with a DevSecOps approach for developing secure software with an automated calculation of the quantitative value of risks and vulnerabilities.

### 3. Analysis of DevOps and DevSecOps approaches

Consider the approaches of using DevOps and DevSecOps in terms of their impact on risks and vulnerabilities in the Software Development Life Cycle. Analyzing DevOps approaches to software development, we can say that at its core DevOps is a philosophy and practice focused on agility, collaboration, and automation in IT processes and a development team [32].

Automation (and the tools that support it) allows developers to adopt agile practices such as continuous integration, delivery, and deployment [33, 34]. This process allows for collaboration across the entire development pipeline, from the concept and development to deployment and testing [35-39].

If applied correctly, a DevOps culture in a project will reduce the time for web application delivery, testing, and deployment. Due to this, the company saves financial resources, all processes are automated. However, in most cases, managers and developers neglect the security issues of web applications and infrastructure and do not include tests in their testing processes that focus on analyzing the information security of an IT product [40, 41]. This increases the security risks of software development, infrastructure, and the impact of vulnerabilities.

In the case of DevSecOps, it is an attempt to automate the main security tasks by introducing control over these processes at an early stage of DevOps. This approach is significantly different from what was used before DevSecOps - security control was the final process and was carried out at the end of development [42]. The DevSecOps methodology should be implemented at the very beginning of software development, just like the DevOps methodology. Taking into account the basis of the methodology - DevOps, the general approach to development concerning the DevSecOps methods can be displayed as shown in Figure 1 [43].

DevSecOps helps to ensure security, with the ability to automatically detect vulnerabilities and comply with security rules throughout the process [43-45]. Implementation of DevSecOps helps to reduce the three security risks and three aspects of DevSecOps: people, processes, and technology selection due to the following properties:

- Reducing the number of vulnerabilities, malicious code, and other security issues in software under development without slowing down code production and production.
- Reducing the potential impact of vulnerability exploitation throughout the software lifecycle by taking advantage of modern and innovative technologies.
- Eliminating vulnerabilities and preventing their recurrence, for example, by implementing modern testing technologies, improving code development practices, and working with cloud services.
- Reducing conflict between people so that developers, security professionals, and operations staff can find common ground to maintain the speed and agility needed to support the organization's mission.

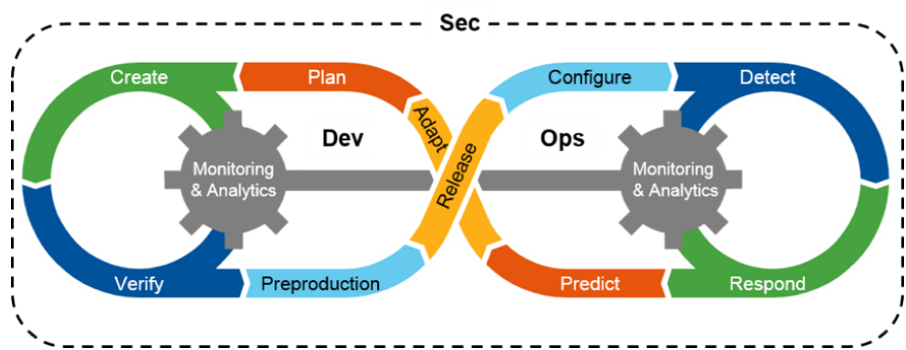

**Figure 1.** Main Stages of the DevSecOps Methodology

The benefits of DevSecOps are easy to highlight - process automation from the very beginning reduces the likelihood of mismanagement and errors that often lead to downtime or open up space for attacks [5, 46-48], which will increase the risks of the organization.

Based on the analysis of all the most popular software development methodologies, the paper will present the main methods of infrastructure deployment, their comparison, and ways of automation for decrease the risks of organization and number of vulnerabilities. Analyzing these methods is extremely important in terms of information security because the proper configuration of the infrastructure and tools that automate certain processes will allow to have a secure infrastructure and fault-tolerant solutions at the very beginning of software development.

**4. Development of a risk management solution with automatic assignment of risk values for new vulnerabilities or incorrect configurations based on their description**

The main parameters when evaluating security risks are data confidentiality, integrity, and availability (CIA). The risk value is determined based on the probability of events threatening these three areas, and their impact. Based on the calculated risk values measures to reduce it should be developed according to the company's security policy. Modeling threats for security risks is focused on all stages of the software development lifecycle including requirements collection, design, development, and testing [54]. Risks and vulnerabilities that interfere with the operation of software systems are used as a basis for assessing security risks. Let us consider an algorithm for the automated calculation of security risks:

1. Creating a database of system components, software, and processes. For example, network components, software components, etc.
2. Creating a database for tracking events in the system to be used in the automated impact analysis.
3. Analysis of threats and vulnerabilities impact based on CIA. To begin with, this impact analysis may be based on expert assessment. The impact analysis of threats and vulnerabilities should include confidentiality, integrity, and availability. International cybersecurity standards show that this approach to impact assessment provides a comprehensive view of security threats and their impact.
4. Modeling possible threats and vulnerabilities that can influence CIA according to components categories. This means creating a list of threats and vulnerabilities that can be selected from the field-specific knowledge or based on the organization's experience, i.e. a process recording of all threats and vulnerabilities of the system, software, processes and system components that led to the realization of threats. Information about threats to software development and operation processes is based on a list of software and hardware vulnerability types from the field databases and data collected during software development and use.
5. Calculating the impact value when a threat is realized based on CIA. This requires comparing experience and events from the organizational database. The impact values of the CIA threat are determined by the company based on their impact on the consumer, which will allow us to get the total significance of the impact.
6. Calculating the probability of threats and vulnerabilities. The probability of emerging can be assessed qualitatively or quantitatively. To calculate the probability of vulnerabilities, it is necessary to systematically record all events related to CIA.
7. Calculating the risk of threats and vulnerabilities. Based on the types of threats, vulnerabilities and risk, field-specific knowledge bases can be used to provide information on CIA controls.
8. Implementation of controls over information security, threats and vulnerabilities. After implementing the controls, it is possible to list the remaining risks and update the controls.

In other words, the system operates in a real time mode, periodically collecting all information and adjusting the system by the risk value and applied controls. If any changes are made to the system structure, components or processes, the algorithm is run again. Figure 2 shows an algorithm for automated risk assessment based on the calculation (modeling) of threats and vulnerabilities.

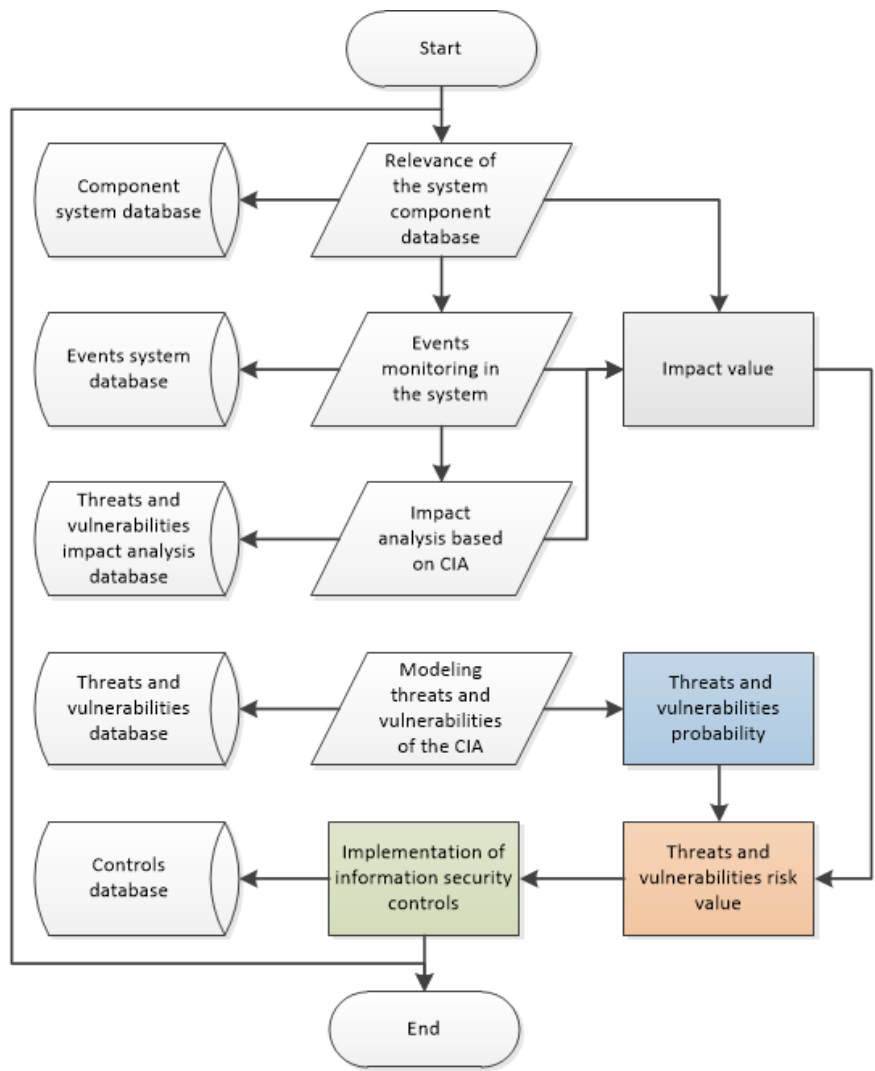

**Figure 2. An algorithm for automated risk assessment based on the calculation (modeling) of threats and vulnerabilities**

This algorithm can be used in any information security management system of the company, as well as in cloud-based systems for software creation and operation.

Any method of calculating risks can be used in this algorithm. As mentioned above, risk values can be qualitative or quantitative [30, 55]. In this paper, we propose an approach based on the modernized FAIR methodology, and the calculated risk value is quantitative [29, 31, 56].

The conventional FAIR methodology is based on the analysis of the frequency of an incident and likely losses from its occurrence [56]. In general, the methodology is divided into four stages: identification of assessment objects, assessment of the frequency of threats, assessment of the probability of potential damage and obtaining and formalizing the risk.

Let us describe each of the stages in more detail.

1. The stage of identification of assessment objects: assets and threats that can be applied to a specific information system are assessed.
2. The stage of assessing the frequency of threats: the methodology is based on a matrix approach, and the frequency is expressed as very high, high, medium, low or very low. Values that correspond to these levels are determined by companies themselves.
3. Calculation of the amount of losses: the methodology uses a scale for converting quantitative values into qualitative ones.

    3.1 To assess the worst-case scenario:
    - Determine the impact of the threat that is likely to result from the worst-case scenario;
    - Estimate the amount of each type of loss associated with the threat;
    - Total the amount of all types of losses.

    3.2 To estimate the likely amount of losses.
4. Formalizing risks on the basis of company's data.

As a result, the calculation of the risk value is reduced to a matrix, where the desired value will be at the intersection of the frequency of events leading to losses and the maximum value of losses of the selected threat.

Thus, the FAIR methodology is a clear and detailed risk assessment process, but its result is not convenient, since the range of risk values can be quite large (the risk value can differ by an order), and companies need quantitative risk values. That is, it is necessary to make a transition from a qualitative assessment of information risks (according to the FAIR methodology) to a quantitative assessment.

This transition involves the following steps:
1. Selecting an asset for which the risk assessment will be performed, e.g. a file containing confidential information located on a computer.
2. Determining:
    - Possible events that lead to a CIA breach, separately for confidentiality, integrity and availability.
    - The impact of threats and vulnerabilities based on CIA.
3. Calculating the probability of threats and vulnerabilities and calculating the impact value when a threat occurs.

To do this, we assume that events that may lead to a breach of confidentiality, integrity or availability of information are independent, i.e. the occurrence of one of these events does not affect the occurrence of another. The realization of any of these events leads to the loss of confidentiality, integrity or availability of information with probability P. Basing on the theorem that the probability of occurrence of at least one of the events, independent in aggregate, is equal to the difference between one and the product of the probabilities of opposite events, we obtain the expression:

$$P = 1 - \prod_{i=1}^{n}(1 - P(A_i)), \qquad (1)$$

where $P(A_i)$ is the probability of the i-th event; n is the number of events that may lead to a CIA violation.

At the same time, the realization of events $A_i$ depends on a number of hypotheses (factors) that are independent of each other within the same event. Let us denote the probabilities of hypotheses realization by $P(H_{ij})$. They are also independent of each other within the same event.

Let $P(A_i|H_{ij})$ be conditionally the probability of the occurrence of event $A_i$ under the j-th hypothesis. Basing on the formula for the total probability and the formula for adding probabilities, we obtain the expression:

$$P(A_i) = \sum_{i=1}^{n} \sum_{j=1}^{m} P(H_{ij}) P(A_i | H_{ij}), \qquad (2)$$

where i is the current event number,

j is the current hypothesis number,

n is the number of relevant events,

m is the number of hypotheses.

Thus, using formulas (1) and (2), we obtain values of the probabilities for confidentiality, integrity and availability breaches separately.

At the same time, the loss value can be calculated using the formula:

$$R = \sum_{i=1}^{n} P_i \cdot E_i, \qquad (3)$$

where Pi is the probability of CIA breach,

Ei is the amount of losses resulting from these events.

Thus, the paper presents an approach to calculating losses based on statistical data.

As an example, we calculated the value of risk that a company may have in case of an information availability breach. As an asset, program code with availability vulnerability was considered. Program code errors were considered as actions that could violate users' CIA and harm the company.

Following the FAIR methodology, risk is calculated as a product of a probable frequency of insured cases and a probable amount of possible losses.

As noted above, risk in terms of an asset is the sum of the products of the probabilities of information confidentiality, integrity, and availability breach by the amount of probable loss from the occurrence of these events.

Thus, the risk value obtained using the proposed method fell within the range of the probable value of losses determined by the standard approach. But, unlike the qualitative assessment of the FAIR methodology, the obtained risk value is quantitative, which allows for more effective forecasting of information security costs.

It should be noted that when calculating according to the developed methodology, it is necessary to take a responsible approach to the process of identifying threats and their causes, and to take into account the company's and analytical agencies' statistical data. If necessary, it is recommended to use expert services.

## 5. Results

*5.1 Finding the most relevant cloud provider using DevSecOps approaches*

When it comes to choosing a cloud provider for deploying web applications in a cloud environment, executives face a difficult task. The analysis has shown that when using the DevSecOps approaches when it comes to decision-making methods, the most relevant method is a hierarchy analysis method developed by Thomas Saaty [49]. This method helps company executives who make decisions about choosing a cloud environment to find the most relevant solution that best suits their understanding of the issue.

Mathematical calculations, analysis, and results of deciding upon the most relevant cloud provider using Th. Saaty's method is presented in this section of the paper.

The general idea of Thomas Saaty's method of hierarchy analysis is to decompose the choice problem into simpler components and process a decision maker's judgments [49]. As a result, the relative importance of the studied alternatives is determined by all the criteria in the hierarchy.

Suppose a company wants to develop and implement a custom web application that will be hosted in the cloud. The company's executives want to have:

- A fault-tolerant cluster and high availability of the application to ensure the availability of information;
- Encryption of traffic between servers and DBMS, as well as between a user and server to ensure integrity;
- Encryption of the logical disk on a server, databases, and relevant tables to ensure confidentiality;
- Financial relevance, i.e. traffic and hosting prices feasibility in the cloud environment;
- The ability to introduce additional services to improve the level of information security in a cloud;
- A wide range of regions to reduce package delivery time;
- The ability to quickly and efficiently launch a new similar infrastructure for different teams, environments, tests, etc.

Taking into account these customer's preferences, let's look at 5 main criteria for choosing a cloud environment.
1. Traffic costs (incoming and outgoing traffic).
2. A number of available regions for application hosting.
3. Cluster costs.
4. A number of services that help to increase the level of software security.
5. The speed of new infrastructure deployment.

To select the most suitable cloud provider for these criteria, it is recommended to use Th. Saaty's hierarchy analysis method. Next, the paper will present calculations for the three most popular cloud providers: Amazon Web Services, Microsoft Azure and Google Cloud Provider.

Supposing a company is not interested in using Kubernetes orchestration as it considers this product to be expensive and difficult to set up and use. The use of Kubernetes should be a well-balanced decision.

So, let a client's future cluster be characterized by the following components:
- Two servers with 2 virtual processors (vCPU) and 4 GB of RAM;
- One database server with 2 virtual processors (vcpu) and 2 GB of RAM;
- One load balancer that will accept requests and redirect them to servers located logically behind it;
- Region - Germany, as it is the closest region to Ukraine and it refers to all three options under study.

The deployment time for a new infrastructure can be considered the operation time of the software that uses the Infrastructure as Code (IaC) approach - Terraform v.0.12. This is an open source software tool created by HashiCorp. This component is independent of a cloud provider, so it is suitable for quick and efficient creation of a new cluster by all popular cloud providers.

For ten consecutive days, experiments were conducted on scaling up the infrastructure with Terraform software for the three analyzed cloud providers in the Germany region [50]. The infrastructure consisted of the cluster components defined above.

It turned out that the average deployment time of a cloud infrastructure is equal 6.3 minutes for AWS; 6.5 minutes for Azure; 7 minutes for GCP.

Using the open source data, we analyzed cluster and traffic costs, as well as a number of publicly available regions and a number of services that increase the level of application security.

Considering that the desired web application is new and still unpopular, we analyzed traffic costs from the first packages up to 10 TB. All the cloud providers analyzed in this section offer lower rates after 10 TB.

The number of regions is given for the 2023 year. Since cloud providers are developing at a tremendous speed, the number of regions is sure to grow over time.

So, all the data concerning the customer's main criteria is shown in Table 1.

Table 1. Data on the main analysis criteria

| Criteria | AWS | Azure | GCP |
| --- | --- | --- | --- |
| Traffic costs (per GB) | 0.085 | 0.0875 | 0.11 |
| Number of regions | 25 | 60 | 24 |
| Cluster costs (per month) | $350.37 | $513.89 | $365.21 |
| Number of security services | 33 | 30 | 26 |
| Deployment time (min) | 6.3 | 6.5 | 7 |

To decide upon the most relevant cloud provider for a company, it is recommended to use Thomas Saaty's method.

The first step in solving the problem is to build a matrix of pairwise comparisons, which is done on a qualitative scale and then converted into points:
- All the same, don't care = 1;
- Slightly better (worse) = 3 (1/3);
- Better (worse) = 5 (1/5);
- Much better (worse) = 7 (1/7);
- Significantly better (worse) = 9 (1/9).

For intermediate opinions, intermediate scores are used accordingly: 2, 4, 6, 8.

The matrix is formed in accordance with the following ratios:

$$\begin{cases} a_{ij} = 1/a_{ji}, \\ a_{ii} = 1. \end{cases} \quad (4)$$

Next, you need to compare alternatives by criteria.

After building the matrices of comparisons of alternatives, it is necessary to use the method of matrix analysis [51].

For this purpose, first we find the sum of the elements of each column using the formula:

$$S_j = a_{1j} + a_{2j} + ... + a_{nj}. \quad (5)$$

Let's normalize the resulting matrix by dividing all the elements of the matrix by the sum of the elements of the corresponding column using the formula:

$$A_{ij} = \frac{a_{ij}}{S_j} \quad (6)$$

Then we find the average value for each row, the result of which is shown in Table 2. The total sum A(ij) in average must be equal to one.

Table 2. Average value for each element

| $\overline{A_{ij}}$ | Traffic costs | Number of regions | Cluster costs | Number of security services | Deployment time | Average |
|---|---|---|---|---|---|---|
| Traffic costs | 0.045 | 0.016 | 0.025 | 0.076 | 0.115 | 0.055 |
| Number of regions | 0.268 | 0.096 | 0.059 | 0.101 | 0.192 | 0.143 |
| Cluster costs | 0.313 | 0.289 | 0.179 | 0.152 | 0.308 | 0.248 |
| Number of security services | 0.358 | 0.578 | 0.716 | 0.606 | 0.346 | 0.521 |
| Deployment time | 0.015 | 0.019 | 0.022 | 0.067 | 0.038 | 0.032 |

The column we got indicates the "weight" of the criteria in terms of the objective set by the client. This column is called a criteria weight column according to the objective.

Thus, we can draw an intermediate conclusion. From the point of view of meeting the company's objective, the most significant are "Number of security services" with 52.1% and "Cluster costs" with 24.8%. "Traffic costs", "Number of regions" and "Deployment time" have the lowest coefficients, which in total equal 23%.

After that, it is necessary to make similar (using the same formulas) calculations, this time not for the criteria, but for options, i.e. for cloud providers [52].

Now it remains to determine the weight coefficients of alternatives by multiplying the resulting matrices and we obtain the weights of the alternatives in terms of achieving the objective, which are presented in Table 3, i.e. choosing the most relevant cloud provider among the three options:

Table 3. Weights of alternatives according to the objective

| Cloud provider | Value in decimals | Value in percentages |
|---|---|---|
| AWS | 0.509 | 50.9% |
| Azure | 0.336 | 33.6% |
| GCP | 0.155 | 15.5% |

Thus, the Amazon Web Services cloud provider is the most relevant option for the company. Despite the fact that AWS currently has 25 regions, this provider has the shortest time of launching an infrastructure using Terraform software, the lowest prices for using services, and provides the biggest number of useful services that allow to increase the information security level compared to Azure and GCP.

A cloud provider Azure has a huge number of available regions around the world - 60, which is more than AWS and GCP have together. Azure also has a large number of security services, but having a cluster with this provider is currently not cost-effective.

Google Cloud Platform is the youngest provider of all those analyzed, so there is every chance for it to reach the level of AWS and Azure in the future. If we were analyzing a web application that used containerization (e.g. Docker) with the Kubernetes orchestration system, Azure would definitely become the most relevant provider, as Kubernetes and Azure are both Google projects.

It is worth noting that this analysis was an expert one and was based on the requirements we set ourselves. Thus, each company, when planning software development and using Agile methodologies, has the opportunity to find out which cloud provider matches its criteria best using Th. Saaty's hierarchy analysis method [51].

Based on the results of the analysis shown in this section, as well as the main approaches to developing secure software, the paper will further present recommendations for creating a fault-tolerant and secure infrastructure with the AWS cloud provider, as well as methods for improving information security when deploying and implementing web applications.

*5.2. Development of a fault-tolerant cluster in a cloud environment based on DevSecOps technologies, which automates the process of secure data use and vulnerability risk calculation*

According to the analysis of agile methodologies and the results of selecting the most relevant cloud provider for the future web application, this section provides recommendations for rapid infrastructure scale up using a Terraform utility, reviews the main AWS services, including those that provide additional software security, analyzes the Jenkins system used for building projects, testing, vulnerability detection and deployment. Moreover, timely notifications have been set up in case a web application or server becomes unavailable or has problems.

Taking into account six main steps of software development, it all starts with planning and analyzing a client's requirements. Let's assume that a client wants to have a highly available and fault-tolerant cluster in the Amazon Web Services cloud environment, the criteria for which were discussed in the previous section of this paper.

As more and more companies are getting interested in a fast and efficient method of scaling up a new infrastructure, it is recommended to use software that applies Infrastructure as Code (IaC) methods. The most popular IaC utilities include Terraform and the AWS CloudFormation service since the project will use AWS. However, in practice, very few companies use AWS's own service because it has a complex syntax and this service can only build an infrastructure with AWS, which makes it impossible to make hybrid solutions when software can be used by several cloud providers.

Instead, Terraform software is becoming more and more popular. It is a free product by Hashicorp that has gained its reputation among DevOps and DevSecOps professionals for such reasons:

- It is cloud-independent. In a modern data center, a customer may have several different clouds and platforms to support different applications. With Terraform specialists can manage heterogeneous environments with the same workflow by creating a configuration file according to the needs of the project;

- Its files have a .tf extension and can be written in JSON format or using its own syntax - Hashicorp Configuration Language (HCL);
- Terraform creates a state file when you first initialize your project, called terraform.tfstate. Terraform uses this file to create plans and make changes in the infrastructure. Before any operation, Terraform updates this file to synchronize the current state of the real infrastructure. It means that the Terraform state is the source of truth through which configuration changes are measured. If a change is made or a resource is added to the configuration, Terraform compares these changes to the state file to determine which changes lead to the creation or modification of a new resource;
- When creating a new resource, deleting components or making some modifications, Terraform always warns the user about all upcoming changes.

As mentioned above, Terraform creates a file called terraform.tfstate when creating an infrastructure. This file contains all the information about the infrastructure, as well as all the data that is transmitted using Terraform. From the point of view of information security, it is forbidden to store this file in the local environment, i.e. on the server where Terraform is running. The best practice is to use AWS S3 service to store files. This prevents accidental deletion of this file, theft of its data, and allows dozens of DevOps and DevSecOps specialists to work on the same infrastructure.

Let's take a closer look at all the components of the future infrastructure, AWS services, including those that provide information security.

Containerization with Docker allows us to solve problems of unstably running on different OS by abstracting from the host. The application is divided into components by functions, which are individually packaged with dependencies, and then can be deployed on an architecture other than a standard one. This feature simplifies the recovery of individual components and scaling of applications as a whole.

Therefore, it is advisable to use this product for your future project. AWS has its own services for storing and using Docker Images, which are easy to configure and provide reliability and security when properly configured. A vulnerability calculation and risk calculation service that interacts with the AWS vulnerability search service was added to the selected architecture.

Having put together all the ideas about the future infrastructure and all the services of the AWS cloud provider, it is always recommended to create a graphical diagram of a future web application. Thus, Figure 3 shows a detailed diagram of the infrastructure that makes software reliable and highly available at every stage of the production lifecycle, and automatically calculates the organization's vulnerabilities and risks.

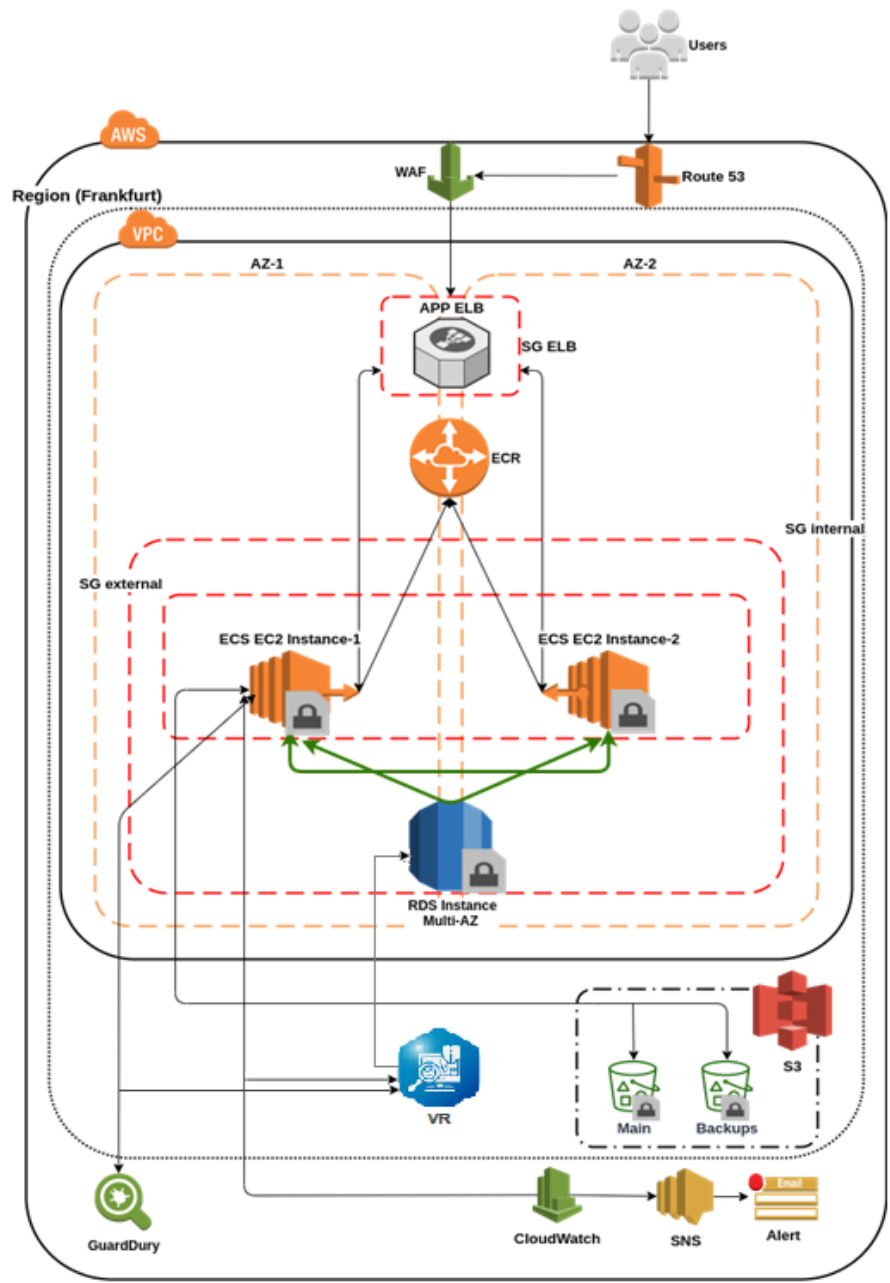

**Figure 3.** A block diagram of a highly available and fault-tolerant infrastructure for a web application in AWS

A detailed description and purpose of each service shown in Figure 3 is provided in [50]. The implementation, and analysis of AWS services used and ways of increasing the level of information security when using them are provided below.

Elastic Compute Cloud (EC2). Select appropriate server characteristics according to the project (t3.medium was used in this project). Encrypt disks (EBS volumes) using your own keys or standard keys from AWS. Create a reliable SSH key using the RSA algorithm (2048 bits and more). Use the latest version of the SSH protocol (v.2). Replace the standard port (22) in configuration files with any other port (>1024). This project used port 37337 instead of 22. Periodically change your private key for a new one and update a public key on the server in the authorized_keys file.

Elastic Container Registry (ECR). Use only private Docker Images, and grant permission to download images only to certain people (DevOps or DevSecOps engineers) and CI/CD systems (e.g. Jenkins or Gitlab-CI).

Elastic Container Service (ECS). Use an Autoscaling Group in the ECS Service component. This will allow you to always have the required number of working servers and launch identical new ones (on condition of a large number of requests), which ensures high reliability and fault tolerance of a web application. In this paper, we used an autoscaling group with parameters min=2, max=4, desirable=2.

Virtual Private Network (VPC). Create your own VPC with at least 2 subnetworks in different availability zones (AZ). Create a separate private subnetwork for a database server. A private subnetwork does not have direct access to the global Internet, so only those servers that are located in the same region and the same VPC will be able to connect to the DBMS directly, as if via a local network. This creates additional protection against intrusion for both regular servers and a database server.

Security Group (SG). Create at least three security groups: for servers created by ECS, for a database server, and a load balancer. Since each component requires open ports to the WAN, it is advisable for information security reasons to open only the necessary ports in security groups, as shown in Figure 4 next to the SG area.

Relational Database Service (RDS). Use RDS in a created private subnetwork (without access to the Internet) and in a security group specific to the database. Encrypt data using your own keys or standard keys from AWS. Create a reliable password for access through a MySQL console that meets 3 out of 4 requirements. Use a multi-az function, which automatically deploys database servers in different availability doses ensuring reliability.

Load Balancer (LB). Upload your own certificate for SSL Termination (HTTPS protocol). Use only the latest versions of TLS - v.1.2 and v.1.3. Add a load balancer to the appropriate security group so that LB can receive only requests coming on ports 80 and 443.

Route53. Set up a redirect if a client has requested an invalid page. For example, our application works at: https://amczi.com, and a client has entered in the search: https://www.amczi.com.This will allow us to have a single central access point to our web application.

Simple Storage Service (S3). Use encryption for objects that are stored in S3 and uploaded there. Create a special user in IAM and grant the user appropriate access only for S3 buckets. Configure ACLs for S3 buckets only for the created user. A web application will be able to use this user's software access, which increases the level of security when storing files and media objects.

CloudWatch and Simple Notification Service (SNS). Enable the collection of extended logs and server data. Based on this data, set up SNS notifications if there are failures in the system or on servers. It is recommended to configure alerts in situations where the servers have high CPU and RAM usage, financial limits, and abnormal actions caused by the GuardDuty service.

Vulnarability and risk assessment Service (VR). VR interacts with RDS to store the database for the automated risk assessment algorithm based on the calculation (modeling) of threats and vulnerabilities. Data from CloudWatch, Simple Notification Service (SNS), and GuardDuty threat detection service are sent to VR to analyze the system status and identify vulnerabilities.

Web Application Firewall (WAF). Create a web access control list (ACL), and select AWS resources for which AWS WAF will check web requests. In my case, these are ECS, RDS, LB, and others. Add rules and groups of rules to be used for filtering web requests. For example, you can specify IP addresses from which requests originate and values in the request that are only used by attackers. For each rule specify whether to block or allow certain web requests. Rules defined within a group of rules have their own actions defined within a group of rules. In this study, we used rules that blocked incoming requests if WAF detected cross-site scripting (XSS) elements, SQL code that could be malicious, and if IP addresses or address ranges from which requests were received reached a certain value over time (3000 requests per 1 hour were configured).

GuardDuty. Enable the service and configure the vulnerability search (AWS GuardDuty Findings). Since the service has three threat levels, you should immediately resolve notifications that have orange and red status (medium and high threat level respectively).

IAM. Create a user with Administrator rights, where software access of this user is required for Terraform implementation. Create an IAM user for the web application to work with S3 buckets. This user should have rights only to work with the S3 service (for example, AmazonS3FullAccess) or create a custom IAM policy for this user that provides flexible security settings for this user.

After analyzing all the services shown in Figure 3 and information above, we can assume that the web application is secure in terms of infrastructure. So, this infrastructure has:

- Encrypted disks on EC2 servers;
- High reliability of working servers due to the use of at least two availability zones. If there is a power failure in one AWS data center, another server located in another AZ will continue to operate and process web requests;
- Sshv.2 access to the servers was performed using secure and large SSH-RSA keys, which was connected on a unique port (37377 instead of 22);
- A permanent number of working servers (2). If the incoming traffic increases and servers reach critical CPU and RAM levels, the system automatically adds up to 4 servers in total;
- A load balancer that receives requests from AWS WAF and automatically distributes them among all working servers. When new servers start up, LB automatically adds them to the distribution;
- An encrypted communication channel between EC2 instances and the RDS server using a pem key that is implemented on the EC2 instance and in the database itself;
- A WAF service, which works according to the groups of rules specified for processing. It filters incoming requests for known attacks on information networks and protects the software from unauthorized access;
- The use of Docker containerization, which allows you to deploy the application on any operating system, provided that the Docker Engine is available;
- Its own virtual network, which ensures reliability and limited access to components located in this network;
- Public subnetworks for EC2 servers and a private subnetwork for the RDS database server, which denies access to all but servers that interact with this database;
- Encrypted S3 buckets that store software backups and media files;
- A group of automatic scaling both horizontally and vertically: when there is not enough server capacity, vertical scaling is used. For example, we used t3.medium instance types. Under heavy loads, an autoscaling group could automatically create new, but more powerful ones, for example, t3.large, which is twice as powerful as the previous ones;
- AWS GuardDuty service, which is used to search for anomalies and problems in the account where a web application is running;
- An automated vulnerability and risk calculation service VR that interacts with data from the organization, CloudWatch, RDS, SNS and AWS GuardDuty;

- Encrypted connection between a client and a web application through the use of HTTPS protocol and our own SSL certificates;
- Timely SNS notifications if our system has problems or metrics on the servers are not satisfactory for us;
- Encrypted data in the database and encrypted communication between EC2 and RDS servers.

The Green-Blue Deployment model using the Jenkins system was used in this study [47]. Jenkins is a system for CI/CD processes (Continuous Integration – CI, Continuous Delivery - CD). In this work we implemented SonarQube in Jenkins Pipeline to find dangerous elements and vulnerabilities [47].

Thus, a Jenkins server was launched and correctly configured, which received push triggers from the GitHub version control system, where program files of the web application were stored. A SonarQube plugin was installed and configured to check the code for vulnerabilities. A Jenkins Pipeline was written in the Groovy programming language and a Jenkins Job was created to perform the CI/CD process. The algorithm was as follows:

- Programmers make new changes to the code and commit to the github repository (git push);
- Jenkins uses webhooks to constantly monitor the repository for new commits in a specific branch (git branch);
- When a new version of the code is released, Jenkins downloads the latest release (git pull) and tests it for code cleanliness and vulnerability using sonarscanner for Jenkins while calculating the risk CIA to the organization;
- If the result is positive, Jenkins builds a Docker Image with the latest software version, sends it to ECR, and then Green-Blue Deployment is performed using the AWS codepipeline service;
- After using the AWS CodePipeline service, we have the latest version of the web application without downtime, which ensures absolute availability of our web application.

As part of the study, the authors conducted an experiment in a configured cluster to automatically calculate the risk using the algorithm proposed in Section 4. Confidentiality (violations in the process of testing the software implementation of a web application in terms of user authentication), integrity (implementation of traffic encryption from the software developer to the cloud service), and availability (implementation of a DDoS attack) were considered as assets. The actions that could harm the company were considered to be those of an attacker aimed at violating the confidentiality, integrity and availability of information in the three areas mentioned above. The results of calculating the probability of a breach of information security and calculating the risk value are shown in the table. The risk value is given in conventional units.

**Table 4**. Results of the calculation of the risk and probability of violation of the CIA

| Value | Confidentiality | Integrity | Availability |
|---|---|---|---|
| Probability | 0.77 | 0.68 | 0.81 |
| Risk assessment | 4184.6 | 3819.3 | 4475.5 |

Thus, with the correct use of tools such as Terraform and Jenkins with the necessary plugins, it is possible to automate business processes and have a reliable and secure infrastructure with the AWS cloud provider. The use of DevSecOps practices ensures the automation of each step of software development, checking program code for open vulnerabilities, calculation of CIA risk in the organization, and solving issues related to the creation of highly loaded information systems in cloud environments.

## 6. Conclusion

The paper addresses the issues of creating a secure and reliable infrastructure in the cloud environment, automatic calculation of vulnerabilities and CIA risk in the organization, as well as methods for deploying web applications by introducing the DevSecOps approaches into the software development process.

The expediency of implementing and using DevSecOps approaches in a software development team has been shown.

The article analyzes modern software development methodologies, provides a comparison of public cloud providers and estimates the most relevant cloud environment using the hierarchy analysis method - a decision-making technology based on mathematical calculations and the use of Thomas Saaty's pairwise comparison method.

All the services of the three today's most popular cloud providers have been analyzed for features increasing the information security level when working with them.

A thorough analysis showed that Amazon Web Services is currently the most relevant cloud provider. This provider has a large number of services that help to increase the level of information security, has components for creating a reliable infrastructure solution, and is financially feasible for companies that have running software in the US and Europe.

Taking into account all the detailed analyses in the paper, a recommended infrastructure cluster scheme was created to implement a web application in the Amazon Web Services cloud environment in which was implemented a service for automatic calculation of vulnerabilities and CIA risk in the organization. This scheme ensures software availability through the use of clustering, product reliability through proven AWS services, and information integrity through data and packet encryption on servers and in information tunnels between them. Instant notifications were set up using the AWS SNS service, which allows you to always be informed in case of abnormal service behavior and when problems with servers are detected. An autoscaling group was set up to provide flexibility and reliability with a large volume of incoming and outgoing traffic. . The software implementation of the automated risk assessment algorithm was set up. Any change in the settings, functioning, and processes in software development and in the organization automatically recalculates the vulnerabilities and risks of the CIA.

The paper develops an algorithm for automated risk assessment based on the calculation (modeling) of threats and vulnerabilities of cloud infrastructure. The proposed system operates in real time, periodically collecting all information and adjusting the system in accordance with the risk and applied controls. When any changes are made to the system structure, components, or processes, the algorithm is started from the beginning. The risk value obtained using the proposed algorithm is quantitative, in contrast to the qualitative assessment of the FAIR methodology, which makes it possible to effectively predict information security costs in software development.

On the basis of the infrastructure diagram, program code in HCL programming language was written to create a cluster with all the necessary services using a Terraform utility. Applying the best practices of the DevSecOps methodology, a Groovy file was created for CI/CD processes with the help of the Jenkins server.

The SonarQube plugin was installed, configured and analyzed to identify vulnerabilities in the program code. The software implementation of the automated risk assessment algorithm was set up.

The results of the research should be used in all IT companies that need to have a secure, fault-tolerant and highly loaded infrastructure in cloud environments, as well as a CI/CD process that will allow managers or responsible people to check their own program code for vulnerabilities at the earliest stages of software development and calculate the risks of the organization and cloud infrastructure.

**Funding:** This research was funded by the Deanship of Scientific Research at Prince Sattam Bin Abdulaziz University under the research project No (IF2/PSAU/2022/01/22780)

**Acknowledgments:** The authors extend their appreciation to the Deputyship for Research & Innovation, Ministry of Education in Saudi Arabia for funding this research work through the project number (IF2/PSAU/2022/01/22780)

**Conflicts of Interest:** The authors declare no conflicts of interest.

**References**

1. Xiong, K. *Resource Optimization and Security for Cloud Services.* John Wiley & Sons Inc.: NY, USA, 2014; 194 p. https://doi.org/10.1002/9781118898598
2. Arshad, M.: Saeed, M.; Rahman, A. U.; Mohammed M. A.; Abdulkareem, K. H.; Alghawli, A. S.; Al-qaness, M. A. A robust algorithmic cum integrated approach of interval-valued fuzzy hypersoft set and OOPCS for real estate pursuit. *PeerJ Computer Science* **2023**, *9*, e1423. https://doi.org/10.7717/peerj-cs.1423
3. Martovytskyi, V.; Argunov, V.; Ruban, I.; Romanenkov, Y. Developing a risk management approach based on reinforcement training in the formation of an investment portfolio. *Eastern-European Journal of Enterprise Technologies* **2023**, *2(3(122))*, 106-116. https://doi.org/10.15587/1729-4061.2023.277997
4. Mulesa, O.; Geche, F.; Batyuk A.; Myronyuk, I. Using A Systematic Approach in the Process of the Assessment Problem Analysis of the Staff Capacity Within the Health Care Institution. In proceedings of 2018 IEEE 13th International Scientific and Technical Conference on Computer Sciences and Information Technologies (CSIT), Lviv, Ukraine, September 11-14, 2018, pp. 177-180. https://doi.org/10.1109/STC-CSIT.2018.8526749
5. Swaraj, N. *Accelerating DevSecOps on AWS. Create Secure CI/CD Pipelines Using Chaos and AIOps*; Packt Publishing: Birmingham, UK, 2022, 520 p.
6. Mulesa, O.; Melnyk, O.; Horvat, P.; Tokar, M.; Peresoliak, M.; Kumar, H. Modeling of Decision-Making Processes in the Service Management System. In proceedings of 2023 IEEE 18th International Conference on Computer Science and Information Technologies (CSIT), Lviv, Ukraine, , October 19-21, 2023, pp. 1-4. https://doi.org/10.1109/CSIT61576.2023.10324217
7. Sood, A, K. *Empirical Cloud Security, Practical Intelligence to Evaluate Risks and Attacks*, 2nd ed., Mercury Learning and Information: Berlin/Boston, Germany/USA, 2023, 462 p.
8. Kirichenko, L.; Radivilova, T.; Bulakh, V. Machine Learning in Classification Time Series with Fractal Properties. *Data* **2019**, 4(1):5. https://doi.org/10.3390/data4010005
9. Daradkeh, Y. I.; Kirichenko, L.; Radivilova, T. Development of QoS methods in the information networks with fractal traffic. *Int. J. of Electronics and Telecommunications* **2018**, *64*(1), 27-32. https://doi.org/10.24425/118142
10. Kirichenko, L.; Zinchenko, P.; Radivilova, T. (2021). Classification of Time Realizations Using Machine Learning Recognition of Recurrence Plots. In: Babichev, S., Lytvynenko, V., Wójcik, W., Vyshemyrskaya, S. (eds) Lecture Notes in Computational Intelligence and Decision Making. ISDMCI 2020. *Advances in Intelligent Systems and Computing*, vol 1246. Springer, Cham. https://doi.org/10.1007/978-3-030-54215-3_44
11. Kirichenko, L.; Radivilova, T. Analyzes of the distributed system load with multifractal input data flows. In proceedings of 2017 14th International Conference The Experience of Designing and Application of CAD Systems in Microelectronics (CADSM), Lviv, Ukraine, February 21-25, 2017, pp. 260-264. https://doi.org/10.1109/CADSM.2017.7916130.

12. Lemeshko, O.; Yeremenko, O.; Mersni A.; Yevdokymenko, M. Resilience Aware Traffic Engineering FHRP Solution. In proceedings of 2021 IEEE International Conference on Information and Telecommunication Technologies and Radio Electronics (UkrMiCo), Odesa, Ukraine, 29 November - 3 December 2021, pp. 1-5. Https://doi.org/10.1109/UkrMiCo52950.2021.9716677.
13. Lemeshko, O.; Yevdokymenko, M.; Yeremenko, O.; Kuzminykh, I. Cyber Resilience and Fault Tolerance of Artificial Intelligence Systems: EU Standards, Guidelines, and Reports. In proceedings of CPITS Kyiv, Ukraine, July 7, 2020, pp. 99-108.
14. Akbar, M. A.; Smolander, K.; Mahmood, S.; Alsanad, A. Toward successful DevSecOps in software development organizations: A decision-making framework. *Information and Software Technology* **2022**, 147, 106894. https://doi.org/10.1016/j.infsof.2022.106894.
15. Rajapakse, R. N.; Zahedi, M.; Babar, M. A.; Shen, H. Challenges and solutions when adopting DevSecOps: A systematic review. Information and software technology **2022**, 141, 106700.
16. Acheampong, R.; Balan, T. C.; Popovici, D.-M.; Rekeraho, A. Security Scenarios Automation and Deployment in Virtual Environment using Ansible. In proceedings of 2022 14th International Conference on Communications (COMM), Bucharest, Romania, 2022, pp. 1-7, doi: 10.1109/COMM54429.2022.9817150.
17. Zhou, X. (57194062816); Mao, R; Zhang, H.; Dai, Q.; Huang, H.; Shen, H.; Li, J.; Rong, G. Revisit security in the era of DevOps: An evidence-based inquiry into DevSecOps industry. *IET Software* **2023**, 17 (4), pp. 435 – 454. https://doi.org/10.1049/sfw2.12132
18. Díaz, J.; Pérez, J. E.; Lopez-Peña, M. A.; Mena G. A.; Yague, A. Self-Service Cybersecurity Monitoring as Enabler for DevSecOps. in *IEEE Access* **2019**, vol. 7, pp. 100283-100295. https://doi.org/10.1109/ACCESS.2019.2930000.
19. Ramaj, X. A DevSecOps-enabled Framework for Risk Management of Critical Infrastructures. In p0roceedings of 2022 IEEE/ACM 44th International Conference on Software Engineering: Companion (ICSE-Companion), Pittsburgh, PA, USA, 22-24 May 2022, pp. 242-244. https://doi.org/10.1145/3510454.3517053.
20. Díaz O.; Munoz, M. Reinforcing DevOps approach with security and risk management: An experience of implementing it in a data center of a mexican organization. In proceedings of 2017 6th International Conference on Software Process Improvement (CIMPS), Zacatecas, Mexico, 18-20 Oct. 2017, pp. 1-7. https://doi.org/10.1109/CIMPS.2017.8169957.
21. Sharma, A.; Singh, U. K. Modelling of smart risk assessment approach for cloud computing environment using AI & supervised machine learning algorithms. *Global Transitions Proceedings* **2022**, Vol. 3, Iss.1, pp. 243-250. https://doi.org/10.1016/j.gltp.2022.03.030.
22. Maniah; Soewito, B.; Lumban F. G.; Abdurachman E. A systematic literature Review: Risk analysis in cloud migration. *J. of King Saud University – Comp. and Information Sciences* **2022**, Vol.34(6B), pp. 3111-3120. https://doi.org/10.1016/j.jksuci.2021.01.008.
23. Curcic, D.; Gupta R.; Narayan, K.; Somasamudram, P.R.; Sarukkai, S. Cloud service usage risk assessment. United State Patent, No .: US 11,521,147 B2, Dec . 6 , 2022.
24. Ahmad, A.; Norziana, J. A Comprehensive Review of Existing Risk Assessment Models in Cloud Computing. *J. of Physics: Conference Series* **2018**, 1018 (1), art. no. 012004. https://doi.org/10.1088/1742-6596/1018/1/012004.
25. Wen-Lin, S; Ying-Han, T.; Yu-Lun H. HiRAM: A hierarchical risk assessment model and its implementation for an industrial Internet of Things in the cloud. *Software Testing Verification and Reliability* **2023**, 33 (5), art. no. e1847. https://doi.org/10.1002/stvr.1847
26. Irsheid, A.; Murad A.; AlNajdawi, m.; Qusef, A. Information security risk management models for cloud hosted systems: A comparative study. *Procedia Computer Science* **2022**, Vol.204, pp. 205-217. https://doi.org/10.1016/j.procs.2022.08.025.
27. T. Weil, "Risk Assessment Methods for Cloud Computing Platforms," in *IT Professional* **2020**, vol. 22, no. 1, pp. 63-66. https:doi.org/10.1109/MITP.2019.2956257
28. Hsu, T. -C.; Pan, Y. -S.; Wu, J. -C.; Liu, Y. -Z. An Approach for Evaluation of Cloud Outage Risk based on FAIR Model. In proceedings of 2023 International Conference on Engineering Management of Communication and Technology *(EMCTECH)*, Vienna, Austria, 16-18 Oct. 2023, pp. 1-6. https://doi.org/10.1109/EM-CTECH58502.2023.10296935.
29. Bland, S. Using FAIR and NIST CSF for Security Risk Management. May 18, 2021. Available online: https://securityintelligence.com/posts/using-fair-and-nist-csf-security-risk-management/ (accessed on 24.Aug. 2023).
30. Whelan, C. How NIST CSF Risk Assessments and the FAIR Risk Model Are Complementary. Available online: https://www.risklens.com/resource-center/blog/how-nist-csf-and-the-fair-risk-model-are-complementary (accessed on 18 Aug.2023).
31. Copeland, J. B. NIST Maps FAIR to the CSF - Big Step Forward in Acceptance of Cyber Risk Quantification. Available online: https://www.fairinstitute.org/blog/nist-maps-fair-to-the-csf-big-step-forward-in-acceptance-of-cyber-risk-quantification (accessed on 4 Juune 2023)
32. Wilson, G. *DevSecOps: A Leader's Guide to Producing Secure Software Without By. Compromising Flow, Feedback and Continuous Improvement*. Rethink Press: Norfolk, GB, 2020, 280 p.
33. Maryland D. *DevSecOps Bootstrap from Cloud. One-Click Launch of a Cloud DevSecOps Solution*. Independently Published, 2020, 24 p.
34. Sobchuk, V.; Barabash, O.; Musienko, A.; Svynchuk, O. (2021). Adaptive accumulation and diagnostic information systems of enterprises in energy and industry sectors. *In E3S Web of Conferences* **2021,** Vol. 250, p. 08002. EDP Sciences.
35. Ruban, I. V.; Martovytskyi, V. O.; Kovalenko, A. A.; Lukova-Chuiko, N. V. Identification in Informative Systems on the Basis of Users' Behaviour. In proceedings of 2019 IEEE 8th International Conference on Advanced Optoelectronics and Lasers (CAOL), Sozopol, Bulgaria, 6-8 Sept. 2019, pp. 574-577. https://doi.org/10.1109/CAOL46282.2019.9019446.
36. Mulder, J. *Modern Enterprise Architecture Using DevSecOps and Cloud-Native in Large Enterprises*. Apress: NY, USA, 2023. 205 p.
37. Evolving Software Processes. Trends and Future Directions. Khan, A.A; Dac-Nhuong, L. Eds. Wiley: UK, 2022, 320 p.
38. Alghawli, A.S. Computer Vision Technology for Fault Detection Systems Using Image Processing. *Computers, Materials & Continua* **2022**, 73(1), 1961-1976. https://doi.org/10.32604/cmc.2022.028990
39. Alghawli, A. S. A.,;Al-khulaidi, A. A.; Nasser, A. A.; AL-Khulaidi, N. A.; Abass, F. A. Application of the Fuzzy Delphi Method to Identify and Prioritize the Social-Health Family Disintegration Indicators in Yemen. *International Journal of Advanced Computer Science and Applications* **2022**, 13(5).
40. Radivilova, T.; Kirichenko, L.; Alghawli, A. S. Entropy Analysis Method for Attacks Detection. In proceedings of 2019 IEEE International Scientific-Practical Conference Problems of Infocommunications, Science and Technology (PIC S&T), Kyiv, Ukraine, 8-11 Oct. 2019, pp. 443-446. https://doi.org/10.1109/PICST47496.2019.9061451.
41. Astapenya, V.; Sokolov, V.; Skladannyi, P.; Zhyltsov, O. Analysis of Ways and Methods of Increasing the Availability of Information in Distributed Information Systems. In proceedings of 2021 IEEE 8th International Conference on Problems of Infocommunications, Science and Technology (PIC S&T), Kharkiv, Ukraine, 12-14 Oct. 2021, pp. 174-178. https://doi.org/10.1109/PICST54195.2021.9772161.
42. Barabash, O.; Sobchuk, V.; Musienko, A.; Laptiev, O.; Bohomia, V.; Kopytko, S. System Analysis and Method of Ensuring Functional Sustainability of the Information System of a Critical Infrastructure Object. In: Zgurovsky, M., Pankratova, N. (eds) *System Analysis and Artificial Intelligence. Studies in Computational Intelligence* **2023**, vol 1107. Springer, Cham. https://doi.org/10.1007/978-3-031-37450-0_11
43. Understanding DevSecOps for Kubernetes. Available online: https://blog.knoldus.com/understanding-devsecops-for-kubernetes/ (accessed on 11 Apr.2023).
44. Wilson, G. *DevSecOps. A Leader's Guide to Producing Secure Software Without Compromising Flow, Feedback and Continuous Improvement.* Rethink Press: Norfolk, GB, 2020, 280 p.
45. Understanding the Differences Between Agile & DevSecOps - from a Business Perspective. Available online: https://tech.gsa.gov/guides/understanding_differences_agile_devsecops/ (Accessed on 16 Mar. 2023).

46. Lemeshko, O.; Yeremenko, O.; Mersni, A.; Gazda, J. Improvement of Confidential Messages Secure Routing over Paths with Intersection in Cyber Resilient Networks. In proceedings of 2022 XXVIII International Conference on Information, Communication and Automation Technologies (ICAT), Sarajevo, Bosnia and Herzegovina, 16-18 June 2022, pp. 1-6. https://doi.org/10.1109/ICAT54566.2022.9811191.
47. Krief, M. Learning DevOps. *A Comprehensive Guide to Accelerating DevOps Culture Adoption with Terraform, Azure DevOps, Kubernetes, and Jenkins*. Packt Publishing: Birmingham, UK, 2022, 560 p.
48. Gorbenko, I. D.; Zamula, A. A.; Semenko, Y. A. Ensemble and correlation properties of cryptographic signals for telecommunication system and network applications. *Telecommunications and Radio Engineering* **2016**, *75*(2). https://doi.org/ 10.1615/TelecomRadEng.v75.i2.60.
49. Saaty, T. L. Analytic Heirarchy Process. Wiley: UK, 2014. https://doi.org/10.1002/9781118445112.stat05310.
50. Cloud Providers and Regions. AWS, Azure and GCP. Available online: https://docs.confluent.io/cloud/current/clusters/regions.html (accessed on 29 Dec. 2022)
51. Magnus, Jan R.; Neudecker, H. *Matrix differential calculus with applications in statistics and econometrics*. John Wiley & Sons:NY, USA, 2019.
52. Blokdyk, G. *DevSecOps Strategy a Complete Guide - 2020 Edition*. Emereo Pty Limited: Queensland, Australia, 2019. 306 p.
53. Datta, D.; Mallick P. K.; Reddy A. V. N.; Mohammed, M. A.; Jaber M. M.; Alghawli, A. S.; Al-qaness, M. A. A. A Hybrid Classification of Imbalanced Hyperspectral Images Using ADASYN and Enhanced Deep Subsampled Multi-Grained Cascaded Forest. *Remote Sensing* **2022**, 14(19), 4853. https://doi.org/10.3390/rs14194853
54. Kipchuk, F.; Sokolov, V.; Skladannyi, P.; Ageyev, D. Assessing Approaches of IT Infrastructure Audit. In proceedings of 2021 IEEE 8th International Conference on Problems of Infocommunications, Science and Technology (PIC S&T), Kharkiv, Ukraine, 12-16 Oct. 2021, pp. 213-217. https://doi.org/10.1109/PICST54195.2021.9772181.
55. Mulesa, O.; Snytyuk, V.; Myronyuk, I. Forming the clusters of labour migrants by the degree of risk of hiv infection. *Eastern-European Journal of Enterprise Technologies* **2016**, Vol. 3, 4(81), pp. 50-55. https://doi.org/10.15587/1729-4061.2016.71203.
56. Dobrynin, I.; Radivilova, T.; Maltseva, N., Ageyev, D. Use of Approaches to the Methodology of Factor Analysis of Information Risks for the Quantitative Assessment of Information Risks Based on the Formation of Cause-And-Effect Links. In proceedings of 2018 International Scientific-Practical Conference Problems of Infocommunications. Science and Technology (PIC S&T), Kharkiv, Ukraine, 9-12 Oct. 2018, pp. 229-232. https://doi.org/10.1109/INFOCOMMST.2018.8632022.